\documentclass {article}
\usepackage{amsfonts}
\usepackage{epsfig}
\def\bbbone{{\mathchoice {\rm 1\mskip-4mu l} {\rm 1\mskip-4mu l}
{\rm 1\mskip-4.5mu l} {\rm 1\mskip-5mu l}}}

\begin{document}

\title{Doubly Special Relativity: A Kinematics of Quantum Gravity?}

\author{Jerzy Kowalski-Glikman\footnote{Research  partially supported
by the    KBN grant 5PO3B05620.}\\Institute for Theoretical
Physics\\ University of Wroc\l{}aw\\ Pl.\ Maxa Borna 9\\
Pl--50-204 Wroc\l{}aw, Poland}

\maketitle

\begin{abstract}
  I give a brief review on motivations, basic postulates, and recent developments in Doubly Special Relativity Theory.
\end{abstract}

\section{Motivations}

There is a consensus in the high energy physics community that the Holy Grail of the theoretical physics, the Quantum Theory of Gravity, once formulated, will change drastically our understanding of space, time and physical processes. Unfortunately, the complete theory is still not known, but we have some indications, stemming from both string theory and loop quantum gravity, of at least one robust prediction of Quantum Gravity. Namely, we believe that space-time structure becomes ``quantum'' at the length scale of order of the Planck scale $\lambda_{Pl} \sim 10^{-33} $ cm. However, as observed by Amelino-Camelia \cite{gac} the existence of such a scale leads immediately to a serious problem. One can imagine physical processes taking place in the regime close to the Planck scale in which gravitational effects can be negligible, in other words very high energetic processes for which the space-time symmetries might be still described by the Standard Lorentz algebra of Special Relativity. However, in such a regime it is hard to understand how it comes that the Planck length, if regarded as a fundamental length and not as a coupling constant is to be observer-independent. This suggests that perhaps relativistic kinematics is being deformed at the Planck scale, so as to incorporate the observer-independent length (or, mass, or time) scale. Doubly Special Relativity is an example of such deformation\footnote{Similar ideas has been spelled out in \cite{ahluwalia}.}.

Moreover in both inflationary cosmology \cite{cosmo} and in black hole physics \cite{bh} one faces the conceptual ``trans-Planckian puzzle'' of ordinary physical quanta being blue shifted up to the Planck energies, which as advocated by many can be solved by assuming deviation from the standard dispersion relation at high energies, and thus deviation from the standard relativistic kinematics, which might be of the form described by Doubly Special Relativity. It should be also stressed that some Doubly Special Relativity models might provide a resolution of observed anomalies in astrophysical data \cite{astro}. Moreover, predictions of the DSR scenario might be testable in forthcoming quantum gravity experiments \cite{qg}.

\section{Postulates}

Having understood the motivations, let us now list and comment the postulates of Doubly Special Relativity. There are three of them

\begin{enumerate}
\item Relativity postulate: the physical processes look the same for all (inertial) observers;
\item There exists a scale of velocity $c$ which is observer-independent; and
\item There exists a scale of mass $\kappa$ (or length) that is observer-independent.
\end{enumerate}

These postulates look quite natural in view of our introductory remarks of the preceding section, and follow very closely the formulation of postulates of Special Relativity. However, to be fair, one should mention some yet unresolved difficulties with this formulation of the basis of the DSR theory.

First of all, contrary to the Special Relativistic case, in DSR we are lacking operational definition of an inertial observer. The reason is that, contrary to space-time formulation of Spacial Relativity, the DSR theory has been always formulated in the energy-momentum space, and an extension of the DSR to space-time is by no means obvious and unique. This means that one cannot make use of the well known construction of inertial observers in terms of rulers and clocks, whose readings are synchronized with the help of light signals.

Second, even though one postulates the existence of observer-independent scales of velocity and mass,  there is an open problem as to whether there exist physical probes moving with velocity $c$. The problem is that the DSR theory seems to predict that velocity of massless particles is wavelength dependent. One should note however that this conclusion follows from rather naive definition of velocity $v = \partial E/\partial | \vec{P} |$, where $E = E(\vec{P})$ is defined by the second Casimir of the DSR algebra (the problem of definition of velocities in DSR theories has been recently considered in \cite{luknowv}, \cite{rs}, \cite{granik}, \cite{THMT}.) It is clear therefore that the velocity of a probe depends on  momentum carried by it, and of course the velocity equals velocity of light $c$ in the limit when the momentum is small compared to $\kappa c$. However this means that the probe moving with speed of light must have low momentum and therefore large wavelength (though, of course there is no reason to believe that $| \vec{P} | \sim 1/\lambda$ for large momenta), and this makes the standard synchronization procedure unapplicable.

The conclusion is therefore that on the level of physical postulates, based on operational definition of basic physical objects present in the theory, the status of the DSR theory is still unclear. For this reason it is therefore useful to try to construct examples of the DSR theory based on some well defined mathematical structure, and only after having understood them to try to return to the problem of formulating the postulates in the well defined operational way.

\section{$\kappa$-Poincar\'e and $\kappa$-Minkowski} 

The $\kappa$-Poincar\'e  algebra is an archetype of all DSR constructions. It is a quantum (Hopf) algebra being a deformation of the Poincar\'e algebra of special relativity. This algebra has been proposed first in the paper \cite{lunoruto} (see also \cite{rurev} for early review). However the algebra presented there has been written in the so-called standard basis, which does not satisfy the natural requirement that the action of the Lorentz sector integrates to a group. Only few years later in the paper \cite{maru} the bicrossproduct basis was introduced in which the Lorentz sector was undeformed. 

The $\kappa$-Poincar\'e algebra can be extended to an algebra on the phase space of the system satisfying (a) the Lorentz sector of this algebra is undeformed, (b) the action of rotations  on momenta is classical\footnote{Let us note that the deformation can be associated only with one dimension and it follows from the requirement of rotational symmetry that we choose this direction to be timelike. One should remember however that there exists a different $\kappa$-Poincar\'e theory in which one deforms the algebra along null direction.}, (c) the space-time commutators and the ones between positions and momenta are uniquely defined by the co-product and appropriate pairing, and (d) in the limit when the deformation parameter $\kappa \rightarrow\infty$ the algebra becomes the classical phase space algebra, i.e, the Poincar\'e algebra along with the standard canonical commutational relations between positions and momenta (with the trivial co-algebra sector). For all examples (or, better to say, bases) of such algebra  according to postulates (a) and (b):
$$
[M_i, M_j] = i\, \epsilon_{ijk} M_k, \quad [M_i, N_j] = i\, \epsilon_{ijk} N_k,
$$
\begin{equation}\label{1}
  [N_i, N_j] = -i\, \epsilon_{ijk} M_k.
\end{equation}
and
\begin{equation}\label{2}
  [M_i, p_j] = i\, \epsilon_{ijk} p_k, \quad [M_i, p_0] =0
\end{equation}
hold.
\newline

For the bicrossproduct basis we have in addition 
\begin{equation}\label{3}
   \left[N_{i}, p_{j}\right] = i\,  \delta_{ij}
 \left( {\kappa\over 2} \left(
 1 -e^{-2{p_{0}/ \kappa}}
\right) + {1\over 2\kappa} \vec{p}\,{}^{ 2}\, \right) - i\,
{1\over \kappa} p_{i}p_{j} ,
\end{equation}
and
\begin{equation}\label{4}
  \left[N_{i},p_{0}\right] = i\, p_{i}.
\end{equation}
with the first Casimir equal
\begin{equation}\label{5}
 m^2 = \left(2\kappa \sinh \left(\frac{p_0}{2\kappa}\right)\right)^2 - \vec{p}\,{}^2\, e^{p_0/\kappa}.
\end{equation}
It should be noted in passing the the parameter $m$ above is {\em not} the physical mass defined by equation $\frac{1}{m_{phys}}=\lim_{p\rightarrow0}\frac{1}{p}\frac{dp_0}{dp}$, $p =|\vec{p}|$, in fact, the correct expression for physical mass has the form
$$
{m^2_{phys}}=\frac{\kappa^2}{4}\, \left(1-\left(-\frac{m}{2\kappa} +
\sqrt{\frac{m^2}{4\kappa^2} +1}\right)^4\right)^2.
$$

Let us now turn to the co-algebra sector of the $\kappa$-Poincar\'e algebra in the bicrossproduct basis. For our present purposes it would be only necessary to know the co-product  for the momentum sector. One has
\begin{eqnarray}
\displaystyle
&& \Delta(p_{i}) = p_{i}\otimes \bbbone +
e^{-{p_{0}/ \kappa}} \otimes p_{i}\, ,
\cr\cr
\displaystyle
&& \Delta(p_{0}) = p_{0}\otimes \bbbone +  \bbbone \otimes p_{0}\, ,
\label{6}
\end{eqnarray}
It is worth noticing that the bicrossproduct basis is singled out by the condition that the energy $p_0$ co-commutes.

The co-product is of crucial physical importance, because it makes it possible to construct the space-time sector and the phase space of the theory by a step-by-step procedure.  Putting it another way, any construction of the space-time sector is in a sense equivalent to definition of some energy-momentum co-product, and only the one described by eq.~(\ref{6}) has the virtue that together with the commutational relations (\ref{1}--\ref{4}) it furnishes a Hopf algebra. It should be stressed at this point that had we not have this structure in our possession, we would not be able to go beyond the energy-momentum sector.

The is a general procedure of construction of the space-time
commutator algebra from energy-momentum  co-algebra
which results in the following commutators \cite{maru}, \cite{crossalg}, \cite{juse}:
\begin{equation}\label{9a}
[x_0, x_i] = -\frac{i}\kappa\, x_i,
\end{equation}
\begin{equation}\label{9b}
[p_0, x_0] = i, \quad [p_i, x_j] = -i \, \delta_{ij},
\end{equation}
\begin{equation}\label{9c}
 [p_i, x_0] = -\frac{i}\kappa\, p_i.
\end{equation}
Of course, the algebra (\ref{9a}--\ref{9c}) satisfies the Jacobi identity.

\section{Recent developments}

In this section I would like to describe some developments in the field of Doubly Special Relativity that took place in last few month. This short review is far from being complete as I take liberty to describe mostly my own research.

The event that led to drastic change of our view of DSR theories was the proposal of Maguiejo and Smolin \cite{Magueijo:2002am} of a DSR theory that differ from the one described in the preceding section. This raised a natural questions concerning relations between various DSR theories and some structures that are common for all of them. These questions has been raised first in \cite{juse} and \cite{lunoDSR}, while the complete answer has been presented in \cite{Kowalski-Glikman:2002jr}. It consists of the following:

The universal structures of all DSR theories are

\begin{enumerate}
\item The space-time noncommutative algebra (\ref{9a}) and
\item The algebra of action of boosts on coordinates
\begin{equation}\label{10}
[N_i, x_j] = i \delta_{ij} x_0 - \frac{i}\kappa\, \epsilon_{ijk} M_k, \quad [N_i, x_0] = i x_{i} - \frac{i}\kappa\, N_i.
\end{equation}
\end{enumerate}
These structures can be called $\kappa$-Minkowski space-time.

It is worth noticing that by changing variables $x_i \rightarrow \tilde{x}_i = x_i -  \frac1\kappa
N_i$ one gets another space-time algebra, called Snyder's space-time (cf.~\cite{snyder}) of the form
\begin{enumerate}
\item The space-time noncommutative algebra 
\begin{equation}\label{11}
 [x_0, \tilde{x}_i] = - i\ell^2 \, N_i, \quad [\tilde{x}_i, \tilde{x}_j] =  i\ell^2 \,\epsilon_{ijk} M_k.
\end{equation}
with $\ell = 1/\kappa$ and
\item The algebra of action of boosts on coordinates
\begin{equation}\label{12}
[N_i, \tilde{x}_j] = i x_0, \quad [N_i, x_0]= i \tilde{x}_j,
\end{equation}
\end{enumerate}

The universality of these structures can be easily understood if one turns to the geometric picture underlying the DSR theories developed in \cite{Kowalski-Glikman:2002ft}. In this picture the manifold of momenta is de Sitter space defined by equation
\begin{equation}\label{13}
 -\eta_0^2 + \eta_1^2+ \eta_2^2+ \eta_3^2+ \eta_4^2 =\kappa^2,
\end{equation}
and a DSR theory is nothing but a coordinate system on this space\footnote{It has been shown in \cite{Kowalski-Glikman:2002jr} that such geometrical interpretation holds for all DSR theories.}. In particular, for the bicrossproduct basis described in preceding section we have relations
\begin{eqnarray}
{\eta_0} &=& -\kappa\, \sinh \frac{p_0}\kappa - \frac{\vec{p}\,{}^2}{2\kappa}\,
e^{  \frac{p_0}\kappa} \nonumber\\
\eta_i &=&  - p_i \, e^{  \frac{p_0}\kappa} \nonumber\\
{\eta_4} &=&  \kappa\, \cosh \frac{p_0}\kappa  - \frac{\vec{p}\,{}^2}{2\kappa} \,
e^{  \frac{p_0}\kappa}   \label{14}
\end{eqnarray}
From these perspective the universality of $\kappa$ Minkowski space-time is easy to understand. One notes that the algebra (\ref{9a}), (\ref{10})  is nothing but the $SO(4,1)$ Lie algebra with Lorentz
generators belonging to its $SO(3,1)$ Lie subalgebra.  Let us recall now that both Lorentz
generators and positions can be interpreted as symmetry
generators, acting on the space of momenta as ``rotations'' and
``translations'', respectively. But then it follows that the space
of momenta can be identified with (a subspace of) the group
quotient space $so(4,1)/so(3,1)$ which is nothing but the de
Sitter space. Moreover, it is clear that
 even though  on the (momentum) de Sitter space one can introduce arbitrary
 coordinates (each corresponding to a particular DSR theory), the form of symmetries
 of this space does not, of course depend on the form of the coordinate system. In other words the momentum sectors of various DSR
 theories are in one to one correspondence with differential structures that can be built on de Sitter
 space, while the structure of the positions/boosts/rotations, being related to the symmetries of this
 space is, clearly, diffeomorphism-invariant.

The emergence of Snyder's non-commutative space-time is also clear from this perspective. The symmetry algebra $so(4,1)$ of de Sitter space can be decomposed into its $so(3,1)$ Lorentz subalgebra and the remaining generators in many ways. The $\kappa$ Minkowski space-time (\ref{9a}), (\ref{10}) corresponds to the Levi decomposition of $so(4,1)$, while the Snyder's non-commutative space-time, to its Cartan decomposition.

The geometric picture of DSR theories can be therefore summarized as follows. 
\begin{enumerate}
\item The momentum sector of any DSR theory corresponds to particular coordinate system on De Sitter space.
\item The Lorentz generators and translations in momentum space identified with positions form the $so(4,1)$ algebra of symmetries of De Sitter space
\item Depending on decomposition of this $so(4,1)$ algebra we have to do either with $\kappa$ Minkowski space-time or with the Snyder's non-commutative space-time.
\end{enumerate}

\section{Open problems}

There is, of course, a lot of open problems in Doubly Special Relativity programme, some of which has been already mentioned. In my opinion the most important one is the lack of operational understanding of this theory in space-time; in particular the notion of observer is rather unclear. Another important problem, which is directly related to experimental tests of the theory is our poor understanding of scattering processes, which is important in context of possible DSR explanation of cosmic rays anomalies. Last but not least, if Doubly Special Relativity is to be a kinematical structure at Planck scale we should learn how to construct dynamics, e.g, a field theory for whose the $\kappa$-Poincar\'e algebra is an algebra of kinematical symmetries.

\end{document}